\documentclass[numberedappendix]{emulateapj}

\shorttitle{Advection of Magnetic Fields in Accretion Disks}
\shortauthors{Rothstein \& Lovelace}

\begin{document}

\submitted{Accepted for publication in The Astrophysical Journal}

\title{Advection of Magnetic Fields in Accretion Disks: Not So Difficult After All}

\author{David M. Rothstein\altaffilmark{1,2} and Richard V. E. Lovelace\altaffilmark{3}}

\altaffiltext{1}{Department of Astronomy,
Cornell University, Ithaca, NY 14853-6801;
droth@astro.cornell.edu}
\altaffiltext{2}{NSF Astronomy
and Astrophysics Postdoctoral Fellow}
\altaffiltext{3}{Departments of Astronomy and Applied and
Engineering Physics,
Cornell University, Ithaca, NY 14853-6801;
RVL1@cornell.edu}

\begin{abstract}

We show that a large-scale, weak magnetic field threading a turbulent accretion disk tends to be advected inward, contrary to previous suggestions that it will be stopped by outward diffusion.  The efficient inward transport is a consequence of the diffuse, magnetically-dominated surface layers of the disk, where the turbulence is suppressed and the conductivity is very high.  This structure arises naturally in three-dimensional simulations of magnetorotationally unstable disks, and we demonstrate here that it can easily support inward advection and compression of a weak field.  The advected field is anchored in the surface layer but penetrates the main body of the disk, where it can generate strong turbulence and produce values of $\alpha$ (i.e., the turbulent stress) large enough to match observational constraints; typical values of the vertical magnetic field merely need to reach a few percent of equipartition for this to occur.  Overall, these results have important implications for models of jet formation which require strong, large-scale magnetic fields to exist over a region of the inner accretion disk.

\end{abstract}


\section{Introduction} \label{sec:intro}

\begin{figure}
\epsscale{0.95} \plotone{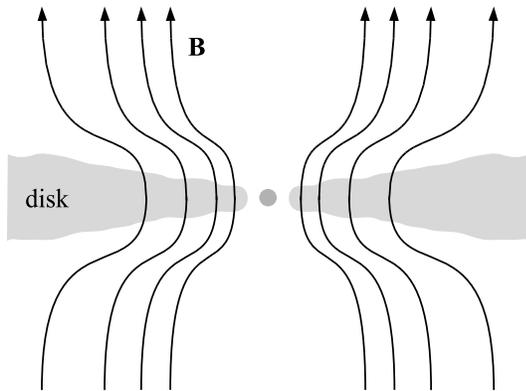} \epsscale{1.0}
\caption{Sketch of the magnetic field threading an accretion disk, showing the increase of the field due to its assumed inward advection with the gas (as proposed in early theoretical models).}
\label{fig:hourglass}
\end{figure}

    Early theoretical work on accretion disks argued that a large-scale magnetic field (of, for example, the interstellar medium) would be dragged inward and greatly compressed by the accreting plasma \citep{Bisnovatyi1974,Bisnovatyi1976,Lovelace1976}.  Figure \ref{fig:hourglass} illustrates this concept by showing a sketch of an ordered magnetic field threading an accretion disk, in which inward advection has caused the magnetic field lines to bunch together into an ``hourglass'' shape.  This was thought to be a simple mechanism for generating dynamically significant fields in the inner disk.

In the present paper, we revisit this issue, building off the recent work of \citet{BKL2007}.  Our motivation for doing so is that in the intervening years, the early theoretical arguments have been challenged.  More detailed models of turbulent disks suggested that a large-scale, weak magnetic field such as that shown in Figure \ref{fig:hourglass} in fact will diffuse outward rapidly \citep*{vanball1989,Lubow1994} if the turbulent magnetic diffusivity and turbulent viscosity are of similar order of magnitude, as they are expected to be \citep{Parker1971,Bisnovatyi1976,Canuto1988}---the turbulence responsible for driving the accretion also leads to enhanced reconnection of the large-scale radial field across the thickness of the disk, thereby causing the vertical field to diffuse away.  This cast doubt on the idea that weak fields could be dragged inward and compressed by advection.  At the same time, it was known that the angular momentum loss to magnetohydrodynamic (MHD) outflows from a disk threaded by a sufficiently {\it strong} large-scale field could more than offset the outward diffusion and lead to a rapid, implosive increase of the field in the central region of the disk \citep*{Lovelace1994}.  However, it seemed to be the case that growth of a strong magnetic field ``from scratch,'' due to continual advection of a weak field, was impossible in a thin disk.  Although this conclusion has been occasionally challenged \citep[e.g.,][]{OgilvieLivio2001}, it is still generally accepted, which has led to the recent suggestion that special conditions (extremely nonaxisymmetric regions of strong field in an otherwise weakly-magnetized disk) are required for the field to be advected inward \citep{Spruit2005}.

At the same time, recent
three-dimensional MHD simulations
have been performed that allow this
issue to be addressed computationally.
   These simulations resolve the
largest scales of magnetorotational
turbulence and therefore
self-consistently include the turbulent
viscosity and diffusivity
(without having to prescribe their values {\it a priori}).
   Most simulations performed to date
have investigated conditions in which the
accreting matter does not contain
any net magnetic flux and where no magnetic field is supplied at the boundary of the computational domain.
     However, in  one simulation, weak
poloidal flux injected at the
outer boundary was clearly observed to
be dragged into the central region of the disk,
leading to the buildup of a
strong central magnetic field \citep*{Igumenshchev2003}.
A similar process, albeit transient, may occur in simulations without a net magnetic flux; there, radial stretching of locally poloidal field lines in the initial configuration often leads to large-scale poloidal fields and jet structures in the inner disk \citetext{e.g., \citealt{Hirose2004,DeVilliers2005,HawleyKrolik2006}; see also the discussion in \citealt{Igumenshchev2003} and \citealt{McKinney2007} and especially the simulations of \citealt{McKinney2004} and \citealt*{Beckwith2007}, which explore the effect of different initial field geometries on the formation of jets}.
The extent to which any of the advection of magnetic field lines seen in numerical simulations requires
the presence of a thick disk or
nonaxisymmetric conditions is unclear.

In light of these numerical results, we return to the question of inward advection of magnetic fields in this paper, allowing for the possibility that the disk is thin and axisymmetric and asking once again whether advection of a weak field is possible under these conditions.  The mechanisms we discuss here can occur in sufficiently ionized regions of any accretion disk; although they are perhaps most widely applicable to disks around black holes (where the large-scale magnetic field arises entirely within the accreting plasma), they are relevant for disks around many other types of accreting objects as well.

The organization of this paper is as follows.  In \S \ref{sec:advection}, we analyze the advection of a large-scale field in an accretion disk
and point out the importance of
the vertical structure of the disk, which was not
taken into account in most previous studies.  Based on an earlier suggestion \citep{BKL2007}, we show that the thin, highly conducting surface layer of the disk, where turbulence is suppressed, allows a large-scale magnetic field to be advected inward and compressed.  In \S \ref{sec:alpha}, we argue that the resulting magnetic flux through the main body of the disk (due to the large-scale field being advected inward) can produce values of the turbulent $\alpha$ parameter that are in accord with observational data.  This is in contrast with numerical simulations of turbulent disks without a net imposed magnetic flux, which are unable to generate large enough turbulent stress.  Finally, in \S \ref{sec:conditions}, we derive detailed conditions on the field strength, geometry and ionization fraction that are required for the field to be advected inward and show that these are typically weak constraints.  Conclusions of this work are summarized in \S \ref{sec:conclusions}.

\section{Magnetic Field Advection at the Surface of an Accretion Disk} \label{sec:advection}

The evolution of the magnetic field ${\bf B}$ in an accretion disk (averaged
over the short timescales of the turbulence) is assumed to be described by the induction equation,
\begin{equation}
\frac{\partial{\bf B}}{\partial t} = {\bf \nabla} \times \left( {\bf v \times B} - \eta {\bf \nabla \times B} \right) ,
\label{eqn:induction_equation}
\end{equation}
where ${\bf v}$ is the plasma velocity,
$\eta =c^2/(4\pi \sigma)$ is the magnetic
diffusivity, $c$ is the speed of light, and
$\sigma$ is the conductivity.\footnote{Note that in a turbulent disk, there can also be an additional term in equation (\ref{eqn:induction_equation}) that we have not included here, which would represent the contribution of a turbulent dynamo to the growth of the large-scale field.  We ignore this term because we are only interested in the growth of magnetic field due to advection, and therefore any local, dynamo-generated field that may be produced is ``extra'' to that which we discuss in this section.}

We assume a disk with
half-thickness $H \lesssim r$ in cylindrical coordinates.  The main body of
the disk is turbulent, and we take the effective diffusivity to be $\eta \sim \nu$, where $\nu$ is the turbulent viscosity.  The turbulence is widely thought to
be due to the magnetorotational instability
\citep{BalbusHawley1991,BalbusHawley1998,Velikhov1959,Chandrasekhar},
which roughly occurs when the magnetic
energy density is less than the thermal energy density.
We therefore assume a weak magnetic field such that this condition
holds in the main body of the disk.

\begin{figure}
\epsscale{1.15} \plotone{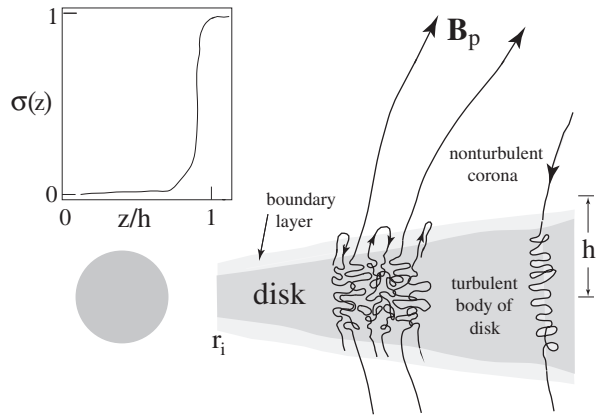} \epsscale{1.0}
\caption{Sketch of the disk and instantaneous poloidal magnetic field lines considered in this work.  The toroidal field component is not shown.  The inset shows a rough illustration of the vertical profile of the conductivity $\sigma \left( z \right)$ in units of the coronal value $\sigma \left( h \right)$.  At the base of the corona ($z=h$), turbulence is suppressed and the conductivity is very high; therefore, if the material in this region advects inward with the main body of the disk, the large-scale magnetic field will be advected inward as well.}
\label{fig:sketch}
\end{figure}

However, the time-averaged magnetic field is not expected to vary strongly across the disk thickness, owing to the buoyancy of the field and the condition ${\bf \nabla \cdot B} = 0$ (in more physical language, the field is not influenced by the vertical gravity that keeps disk material confined near the equatorial plane).  Thus, the mass density of the gas will typically decrease with height $z$ more rapidly than the time-averaged magnetic field strength, and at a height $\sim H$ above the midplane, the magnetic energy density will become strong enough compared to the thermal energy density that turbulence will be suppressed.  The boundary between the turbulent and nonturbulent regions is likely to be ``fuzzy'' owing to the leakage of some magnetic flux through the disk surface \citep*[e.g.,][]{Galeev1979}, but at a certain height, the plasma will become completely nonturbulent.  In this paper, we will use the terms ``base of the nonturbulent region'' and ``surface layer of the disk'' interchangeably; however, it should be noted that we are explicitly defining these regions to be {\it above} the boundary layer and therefore fully a part of the nonturbulent corona (see Figure \ref{fig:sketch}).

This suppression of turbulence
above a weakly-magnetized disk has 
been observed in a variety of MHD simulations
\citep*[e.g.,][]{MillerStone2000,DeVilliers2003,Hirose2004,McKinney2004,Fromang2006},
including those with radiation \citep*{Hirose2006} and even those in which the radiation pressure is comparable to the gas pressure \citep*{Krolik2007}.  However, MHD simulations of {\it fully} radiation-dominant disks \citep{Turner2004} are less clear, and the applicability of our work in this case requires further analysis.  Nonetheless, even above a radiation-dominated region of the disk, we expect that the turbulence will be suppressed in many situations; we discuss this issue further in \S \ref{sec:conditions}.

Figure \ref{fig:sketch} shows a schematic 
drawing of the considered geometry.
   The lack of turbulence at a height $h \sim H$ near
the disk surface causes this layer to become highly
conductive; the diffusivity will decrease from its
turbulent value in the main body of the disk ($\eta \sim \nu \sim 10^{12}$ cm$^{2}$ s$^{-1}$ for typical parameters) to the Spitzer value associated with electrons scattering off of ions, given by
$\eta_S \sim 200 \left( T_{s} / \; {\rm keV} \right)^{-3/2}$
cm$^{2}$ s$^{-1}$, where $T_{s}$ is the surface temperature.\footnote{In actuality, other non-ideal MHD effects, in particular Hall electromotive forces, may be more important than the Spitzer diffusivity, as we show in \S \ref{sec:strength}.  However, the Hall effect usually does not oppose inward advection of the magnetic field, and even if it does, it will not be important unless the field is very weak.}  This suggests that the second term on the right hand side of equation
(\ref{eqn:induction_equation}) can be ignored in
the upper disk layers.
  Specifically, the relative importance of the two 
terms (advection compared to diffusion) at any point in the disk
is determined by the local magnetic Reynolds number
$Re_m =  H u_{r}/\eta$,
where $u_r$ is the local radial speed
and $H$ is the relevant length scale (here we make the reasonable assumption that the time-averaged magnetic field does not vary significantly in the radial direction on length scales shorter than $H$).
We can therefore use the \citet{ShakuraSunyaev1973} disk solution to find that a typical value of the magnetic Reynolds number at the surface of the disk, where turbulence is suppressed, is given by
\begin{equation}
Re_m \sim 10^{8} \; \alpha m^{5/8} \dot{m}^{3/8} \hat{r}^{-5/8} f_{*}^{-5/8} \left( 10^{2} H/r \right)^{3} \mathcal{U}_{s} .
\label{eqn:rem}
\end{equation}
Here, $\alpha \leq 1$ is the dimensionless ``viscosity'' parameter in the main body of the disk (i.e., the magnitude of the turbulent stress divided by the thermal pressure), $m$ is the mass of the central object in solar masses, $\dot{m}$ is the accretion rate in units of the Eddington luminosity divided by the speed of light squared, $\hat{r}$ is the radius in units of the Schwarzschild radius, $f_{*} \lesssim 1$ is a dimensionless function of $\hat{r}$ that depends on the stress at the inner boundary of the disk \citep[e.g.,][]{AgolKrolik2000},
and $\mathcal{U}_{s}$ is the ratio of the radial speed at the disk
surface to that in the main, turbulent body of the disk.  This last term
can be smaller than unity, but not small enough to prevent the conclusion that, typically, $Re_m \gg 1$ at the surface of the disk, and diffusion of the magnetic
field can be neglected.  This is in contrast with 
the main body of the disk, where 
$Re_m \approx H/r$ (assuming the magnetic field
is not strong enough to affect the accretion speed) and
diffusion of the magnetic field therefore
dominates over advection \citetext{\citealt{Lubow1994}; \citealt{Lovelace1994}; \citealt*{Hey1996}}.

We can easily demonstrate that advection 
in the surface layer of the disk
is able to support the overall growth 
of magnetic field.  If we integrate
equation (\ref{eqn:induction_equation}) over a circular
surface $r\leq r_{0}$ that covers the top side of the disk
($z=h$, where $h$ is the height at which turbulence is first suppressed), we can take $\eta \approx 0$, and Stokes' theorem therefore implies that
\begin{equation}
\frac{d \Phi_{p}}{d t} =
r_0\oint d \phi \left( v_{zh} B_{rh} -
v_{rh} B_{zh} \right) |_{r=r_0}~ ,
\label{eqn:magnetic_flux}
\end{equation}
where $\Phi_{p}$ is the poloidal magnetic flux
through this surface and the $h$ subscript indicates that the
quantity is evaluated at $z=h$.  If the right hand side of this
equation has the same sign as
$\Phi_{p}$, the
magnetic flux interior to
radius $r_{0}$ will grow.

Assuming axisymmetry (or, alternatively, treating subsequent quantities as being appropriately averaged over azimuth)
and taking $z>0$ for definiteness,
the condition for magnetic flux growth in equation (\ref{eqn:magnetic_flux})
simplifies to
\begin{equation}
v_{rh} < \left( B_{rh}/B_{zh} \right) v_{zh} ,
\label{eqn:condition}
\end{equation}
provided that magnetic field with the appropriate polarity is available to accrete.

Although turbulent stress cannot contribute directly to the accretion
at $z=h$, coupling between the main, turbulent body of the disk
and the surface (as well as angular momentum loss to a wind or jet)
will tend to produce $v_{rh}<0$.  Also, simulations
indicate that for an MHD outflow or jet ($v_{zh} \geq 0$), the
magnetic field structure has $B_{rh}/B_{zh} \geq 0$ \citep{Ustyugova1999,Ustyugova2000}.  Thus, equation
(\ref{eqn:condition}) is in general likely to be satisfied, a point which we
will discuss more rigorously in \S \ref{sec:conditions}.

In summary, our arguments in this section are a simple consequence of the fact that magnetic fields are sustained by the flow of current, not the flow of mass.  In order to prevent inward advection of magnetic fields, turbulent diffusion must oppose advection throughout the {\it entire} inward-accreting portion of the disk, so that no currents are allowed to accrete inward.  Even a small sliver of nonturbulent (i.e., highly conducting) material that advects inward at the surface layer can support the magnetic field, even though it may only contain a small fraction of the disk's mass.  A related issue was noticed by \citet{OgilvieLivio2001}, who argued (on mathematical grounds) that the relevant radial velocity for magnetic field advection is one that has been weighted by $1/\eta$ and averaged over height.  Here, we present a physical model for the behavior of $\eta$ with height and show that in a disk where the magnetic diffusivity is due to turbulence, the contrast between the diffusivity inside and outside the turbulent region is likely to be so sharp that the condition for magnetic flux growth reduces to equation (\ref{eqn:condition}), which is satisfied in many parts of a typical accretion disk.

\section{Effect of the Advected Magnetic Field on the Turbulent $\alpha$ Parameter} \label{sec:alpha}

Given the apparent ease with which a large-scale magnetic field can advect inward in an accretion disk, it is natural to consider the influence of this magnetic field on the disk dynamics.  In particular, in this section we discuss how the advected magnetic field might be expected to affect the turbulence in the main body of the disk, embodied in the $\alpha$ parameter of \citet{ShakuraSunyaev1973}.  Since the turbulence is thought to be magnetic in nature and in particular due to the magnetorotational instability \citep[MRI;][]{BalbusHawley1991,BalbusHawley1998}, the effect is likely to be a significant one.

\citet*{King2007} have recently pointed out that observationally-determined values of $\alpha$, based primarily on studies of outbursts in dwarf novae and X-ray transients, tend to lie in the range $\sim 0.1-0.4$.  Recalling that $\alpha$ is a measure of the turbulent magnetic stress scaled by the thermal pressure in the main body of the disk, it is clear that a significant amount of turbulent magnetic energy must exist in these accretion disks during the outburst phase.  \citet{King2007} noted a potential puzzle, which is that numerical simulations of the MRI in which the instability is allowed to develop entirely based on a local seed field (i.e., where there is no externally-imposed magnetic flux through the computational region) tend to give saturation values of the stress that are much too small to match the observations, with typical values $\alpha \sim 0.01$ regardless of the strength of the seed field \citetext{\citealt*{Hawley1996,BalbusHawley1998}; and note that \citealt*{Pessah2007} and \citealt{Fromang2007} have shown that even these values may be significant overestimates, due to numerical resolution effects}.

However, in simulations {\it with} an externally-imposed vertical magnetic field, the turbulent stress due to the MRI depends critically on the seed field strength.  In particular, $\alpha$ is found to increase with the net imposed vertical field $B_{z}$ \citep*{Hawley1995,BalbusHawley1998}.  Shearing box simulations suggest a rough empirical relationship $\alpha \sim 2 \pi \left( \beta_{z,ext} \right)^{-1/2}$, where $\beta_{z,ext}$ is the ratio of the thermal pressure in the disk to the magnetic pressure of the externally-imposed vertical field \citetext{this is a simplified version of a more general equation found in \citealt{Pessah2007}, which is based on MRI simulations by \citealt{Sano2004} but also agrees with the earlier results of \citealt{Hawley1995}}.  Thus, values of $\alpha \sim 0.1-0.4$ simply indicate a moderately strong (but still significantly sub-equipartition) vertical field, perhaps with $\beta_{z,ext} \sim 250-4000$.

\citet{King2007} ruled out this mechanism because they believed there was no obvious source for such an externally-imposed field in real accretion disks \citetext{whereas \citealt{Pessah2007} believed it to be possible, but attributed the field to internally-generated MRI fluctuations, rather than an external source}.  However, if large-scale magnetic fields can be advected inward in a disk, and if, furthermore, these fields are maintained by currents flowing in the nonturbulent surface layer of the disk, as we have argued, then MRI simulations with an externally-imposed vertical field are in fact the most relevant for comparing with real accretion disks.  The net vertical field is anchored in the surface layer, and the main, turbulent body of the disk sees this field as a fixed, ``externally-imposed'' seed field.

We therefore conclude that a typical value for the vertical magnetic field in an accretion disk undergoing an outburst, based on observations, is
$\sim 2-6\%$ of the equipartition field strength (i.e., $\beta_{z,ext}$ is $\sim 250-4000$) and that there is no difficultly reconciling the observationally-determined values of $\alpha$ with numerical simulations of the MRI.  Also, as noted by \citet{King2007}, the observationally-determined values of $\alpha$ are weighted averages over the entire accretion disk.  Thus, it is certainly possible that advection could lead to much larger field strengths in a particular region.  The field strengths in quiescent disks are similarly unconstrained.

An interesting effect of the dependence of $\alpha$ on the strength of the large-scale magnetic field is that the viscous and thermal timescales in the disk (which depend inversely on $\alpha$) should change with time, in response to the history of magnetic field advection.  This process may explain some of the wide range of variability on many different timescales seen in accreting black holes, in particular in the bright X-ray binary GRS~1915+105 \citep[e.g.,][]{Belloni2000}, where the various modes of variability seem to repeat in a semi-regular pattern over a period of months to years that has been suggested to be a signature of magnetic processes \citep{Tagger2004}.

A large-scale magnetic field may be expected to have other effects on an accretion disk besides those discussed above.  If the field is strong enough, it can begin to affect the disk dynamics directly (through removal of angular momentum via a wind or jet); we will consider the case where the advected field builds up to dynamically significant values in a future paper.  However, the important point we make in this section is that even when the large-scale field is dynamically weak, it can have a significant effect on the disk dynamics {\it indirectly}, through its influence on the turbulent stress in the main body of the disk.

\section{Detailed Analysis of the Conditions for Magnetic Field Advection} \label{sec:conditions}

In \S \ref{sec:advection}, we showed that advection of a large-scale magnetic field will dominate over diffusion in the nonturbulent surface layer of an accretion disk and that, if equation (\ref{eqn:condition}) is satisfied, the advection can lead to a concentration of magnetic flux in the inner region of the disk.

In this section, we present a more rigorous analysis of the conditions under which equation (\ref{eqn:condition}) is satisfied.  In \S \ref{sec:coupling}, we consider the forces that act between the main body of the disk and the material at the base of the nonturbulent region, which determine whether or not this region is actually a true ``surface layer'' that participates in some way in the accretion flow (i.e., whether or not $v_{rh}<0$ there; clearly, at some height above the disk, the radial velocity may no longer be inward).  In \S \ref{sec:geometry}, we use these results to derive conditions on the magnetic field geometry that allow inward advection of magnetic fields to proceed, and in \S \ref{sec:strength}, we derive more general constraints on the field strength and ionization fraction that are required for a highly conducting nonturbulent surface layer to exist in the first place.  (In general, the above constraints are weak, and whether or not advection can occur is primarily a question of geometry; a vertical or ``dipole-type'' seed field provided in the outer regions of the disk will often be sufficient to advect inward on its own.)  Finally, in \S \ref{sec:outcome}, we consider the ultimate outcome of the advection of a large-scale magnetic field and the types of disks in which it might generally be expected to occur.  Readers not interested in the detailed analysis we present in the rest of this section may wish to skip to the conclusions of this paper in \S \ref{sec:conclusions}.

\subsection{Forces Acting on the Nonturbulent Surface Layer of the Disk} \label{sec:coupling}

For an axisymmetric accretion disk in which the specific angular momentum profile is time-independent (which can be true even in a time-dependent disk if orbits are nearly circular), a general equation for the local radial velocity is
\begin{equation}
v_{r} = \frac{ \displaystyle - 
\left[ \frac{\partial}{\partial r} \left( r^{2} T_{r \phi} \right) +
\frac{\partial}{\partial z} \left( r^{2} T_{\phi z} \right) + r^{2}
\rho v_{z} \frac{\partial v_{\phi}}{\partial z}
\right]}{\displaystyle r \rho \frac{\partial}{\partial r} \left( r
v_{\phi} \right) } .
\label{eqn:vr}
\end{equation}
This expression is obtained by combining the 
conservation of mass and conservation
of angular momentum equations for a magnetized fluid.  
  Here, $\rho$ is the mass 
density and $T_{r \phi}$ and $T_{\phi z}$
are components of the stress tensor, including both large-scale magnetic and small-scale turbulent stresses. If we evaluate this equation at $z=h$ 
(the height in the disk where turbulence is first suppressed) and if the gravitational and centrifugal forces are assumed to balance in this region (i.e., if we assume circular orbits in Newtonian gravity), then to first order in $h/r$,
\begin{equation}
v_{rh} \approx \frac{ \displaystyle 
\frac{\partial}{\partial r} \left( r^{2} B_{r} B_{\phi} \right)_{h} +
\frac{\partial}{\partial z} \left( r^{2} B_{\phi} B_{z}
\right)_{h}}{\displaystyle 2 \pi r \rho_{h} v_{K}} + \frac{3h}{r} v_{zh} ,
\label{eqn:vrh}
\end{equation}
where $v_{K}$ is the Keplerian velocity on the equatorial plane.
The first term represents the effect 
of the stress due to the time-averaged 
magnetic field, and the second
term represents a centrifugal effect 
that drives outflowing material
away from the inner disk.  Competition between these two processes determines the vertical profile of the radial velocity in the nonturbulent region of the disk.

The analysis leading up to equation (\ref{eqn:vrh}) is quite general, and we therefore make liberal use of it in the following sections.  The only exception to its generality is the assumption of circular orbits, which may not be valid if the base of the nonturbulent region occurs at a height $h$ in the disk where the density is so low that radiation pressure or magnetic forces begin to become important in the radial momentum equation.  This is effectively a constraint on the magnetic field strength, and we therefore discuss it further in \S \ref{sec:strength}.  In the meantime, note that equation (\ref{eqn:vr}) does {\it not} assume circular orbits (provided that the disk is stationary) and that the derivation leading from equation (\ref{eqn:vr}) to (\ref{eqn:vrh}) will still be roughly valid provided only that the spatial derivatives of the azimuthal velocity $v_{\phi h}$ in the nonturbulent region are of the same order of magnitude as their Keplerian counterparts.

\subsection{Conditions on the Magnetic Field Geometry} \label{sec:geometry}

A straightforward way to derive a condition on the magnetic field geometry is to combine equations (\ref{eqn:condition}) and (\ref{eqn:vrh}).  If we do this and assume that the stress due to the time-averaged  magnetic field removes angular 
momentum from the disk surface (i.e., attempts to drag the surface layer inward 
with the main body of the disk) {\it in any amount}, then a sufficient condition for growth of magnetic flux in the inner disk is
\begin{equation}
\left( \frac{B_{rh}}{B_{zh}} - \frac{3h}{r} \right)
v_{zh} \gtrsim 0 ,
\label{eqn:mag_growth}
\end{equation}
provided that magnetic field with the appropriate polarity is available to accrete.

When equation (\ref{eqn:mag_growth}) is satisfied, magnetic flux growth can occur through a combination of radial and vertical advection at the surface of the disk.  However, we are primarily interested in radial advection, which is the only {\it sustainable} way in which magnetic field lines anchored in the surface layer can build up flux in the inner disk.  In particular, when equation (\ref{eqn:condition}) is satisfied, we can identify three regimes of interest:
\begin{enumerate}
\item If $v_{rh} < 0 \leq \left( B_{rh} / B_{zh} \right) v_{zh}$, magnetic flux growth occurs through a combination of inward radial advection and vertical advection at the surface of the disk.
\item If $v_{rh} < \left( B_{rh} / B_{zh} \right) v_{zh} < 0$, magnetic flux growth occurs via inward radial advection, even though it is partially opposed by vertical advection.
\item If $0 < v_{rh} < \left( B_{rh} / B_{zh} \right) v_{zh}$, magnetic field is advected radially {\it outward} at the surface of the disk, but magnetic flux growth still occurs in the inner disk because of vertical advection at $z \approx h$.
\end{enumerate}
We are primarily interested in the first two regimes, where magnetic flux growth occurs at least partially due to inward radial advection, and where equation (\ref{eqn:mag_growth}) does not necessarily apply.  Thus, for the rest of this section, we ignore the third regime and derive more stringent conditions that specifically guarantee inward radial advection.

As discussed in \S \ref{sec:advection}, the first regime is likely to be more relevant than the second \citep{Ustyugova1999,Ustyugova2000}, and we therefore consider it now, returning to the second regime at the end of this section.  We thus have $\left( B_{rh} / B_{zh} \right) v_{zh} \geq 0$ and require $v_{rh} < 0$ for inward radial advection.  If we use equation (\ref{eqn:vrh}) and define $H_{B} \equiv \left[ \partial \ln \left( B_{\phi h} B_{zh} \right) / \partial z \right]^{-1}$ as the scale height of the vertical magnetic stress, we find that a sufficient condition is
\begin{equation}
- B_{\phi h} B_{z h} \gtrsim \frac{3}{2} \frac{v_{zh}}{\left< v_{r} \right>} \frac{H_{B}}{H} \left( \frac{\rho_{h}}{\rho_{0}} \frac{h}{r} \right) \frac{\dot{M} \Omega_{K}}{r} ,
\label{eqn:bcondition}
\end{equation}
where $\rho_{0}$ is the mass density on the equatorial plane, $\dot{M}$ is the local mass accretion rate, $\Omega_{K} \equiv v_{K} / r$ is the Keplerian angular velocity, and $\left< v_{r} \right> \equiv \dot{M} / \left( 4 \pi r \rho_{0} H \right)$ is an appropriately height-averaged inward radial velocity in the main body of the disk (i.e., the standard radial velocity in a one-dimensional vertically-integrated disk model).  In interpreting this equation, it is instructive to note that $( \dot{M} \Omega_{K} / r )^{1/2} \approx 1.8 \times 10^{8} \; m^{-1/2} \dot{m}^{1/2} \hat{r}^{-5/4} \; {\rm G}$ is a fiducial field strength, but a maximum one, since $h/r$ and especially $\rho_{h}/\rho_{0}$ can be very small parameters.\footnote{Although we can generally assume $h \sim H$ within a factor of a few, the distinction between those two heights must be retained when evaluating the mass density, and it is important to use the correct value $\rho_{h}$ which appears in equation (\ref{eqn:bcondition}).  This is because the mass density typically falls off very sharply with height, and thus $\rho_{h}$ may be many orders of magnitude smaller than both $\rho_{H}$ and $\rho_{0}$.}

Equation (\ref{eqn:bcondition}) is a ``sufficient'' condition for magnetic field advection in the sense that it makes the conservative assumption that large-scale magnetic stresses in the vertical direction are the only way in which the surface layer can be dragged inward.  In particular, it does not include the large-scale $B_{r} B_{\phi}$ stress at the disk surface, which transports angular momentum radially and also tends to give $v_{rh} < 0$ (although for a thin disk, the vertical stress is usually most important).

The apparent dependence on the field strength in equation (\ref{eqn:bcondition}) disappears when the equation is analyzed more carefully.  In particular, we show in Appendix \ref{appendix:density} that the definition of $h$ as the height in the disk where MRI turbulence is first suppressed allows us to write
\begin{equation}
\frac{\rho_{h}}{\rho_{0}} \sim \frac{\mathcal{B}_{zh}^{2}}{8 \pi p_{0}} ,
\label{eqn:beta_condition}
\end{equation}
where $\mathcal{B}_{zh} \approx \left| B_{zh} \right| + \left( H/r \right) \left| B_{\phi h} \right|$ represents a magnetic field strength that is roughly equal to the time-averaged vertical field $B_{zh}$, and we define $p_{0} \equiv \rho_{0} \left( H \Omega_{K} \right)^{2}$; for a weak field, $p_{0}$ is roughly equal to the thermal pressure on the equatorial plane of the disk.  Combining equations (\ref{eqn:bcondition}) and (\ref{eqn:beta_condition}) and assuming $h \sim H$, we find that a sufficient condition for inward advection of magnetic fields is
\begin{equation}
-\frac{B_{\phi h} B_{z h}}{\mathcal{B}_{zh}^{2}} \gtrsim \frac{H_{B}}{H} \frac{v_{zh}}{v_{K}} ,
\label{eqn:geometric_condition}
\end{equation}
which has no direct dependence on the field strength; as long as the geometry is favorable, arbitrarily weak magnetic fields can provide enough stress to drive inward radial advection at the surface layer of the disk, and the fields will therefore be advected inward and compressed.  The essential physical point is simply that the magnetic field must be strong compared to the gas at $z \approx h$ (in order to suppress the MRI), so it is therefore able to drive accretion at this location, regardless of how weak it is in an absolute sense.\footnote{This statement is independent even of the Spitzer diffusivity; our expression for $v_{rh}$ in equation (\ref{eqn:vrh}) means that we can rewrite the condition $Re_m \gg 1$ (which must be satisfied at the base of the nonturbulent region in order for advection to dominate the Spitzer diffusivity there) as $\left| \left( H/H_{B} \right) B_{\phi h} B_{zh} / \mathcal{B}_{zh}^{2} + v_{zh}/ v_{K} \right| \gg 10^{-11} m^{-5/8} \dot{m}^{-3/8} \hat{r}^{5/8} f_{*}^{-3/8} \left( 10^{2} H / r \right)^{-2}$.  Clearly, this equation will be satisfied in almost any accretion disk provided that equation (\ref{eqn:geometric_condition}) is not pathologically close to an equality.  More important for our purposes, there is no dependence on the magnetic field strength; even when microscopic effects such as Spitzer diffusivity are taken into account, arbitrarily weak fields appear capable of advecting inward along the surface layer of a fully ionized accretion disk (but see \S \ref{sec:strength}).}

If we ignore the sign of $B_{\phi h} B_{zh}$, equation (\ref{eqn:geometric_condition}) is relatively trivial to satisfy.  For example, we can estimate the ratio of vertical magnetic stress to energy density on the left hand side of equation (\ref{eqn:geometric_condition}) that might arise naturally in a disk (i.e., without an externally-imposed seed field) by looking at numerical simulations of the MRI.  We focus on the work of \citet{MillerStone2000}, who studied a vertically-stratified disk in the shearing box approximation and tabulated the properties of the magnetic field in the nonturbulent corona above the disk.  We find typical values of $\left| B_{\phi h} B_{zh} \right| / \mathcal{B}_{zh}^{2} \gtrsim 0.05$ in these simulations.  This should be compared to the right hand side of equation (\ref{eqn:geometric_condition}), whose magnitude is given by $\sim \alpha \left( H/r \right)^{2} \left( H_{B}/H \right) \left| v_{zh} \right| / \left< v_{r} \right>$ when the large-scale field is dynamically weak (or, alternatively, $H_{B}/r$ times the ratio of $v_{zh}$ to the disk sound speed).  This is clearly a very small number if we make the approximation that $H_{B} \lesssim H$ (i.e., that the scale height of the twisted toroidal field can be comparable to or smaller than that of the mass density); the validity of this approximation is discussed in Appendix \ref{appendix:geometry}.  Intuitively, the approximation $H_{B} \lesssim H$ may be thought of as arising from the presence of a voltage source (Keplerian shear) that is applied in the radial direction, with the resulting current confined to flow in a region above $\sim h$ (the height at which the plasma first becomes highly conductive) but below $\sim  {\rm a \; few} \times H$ (the height at which the mass density becomes low enough so that orbits are no longer circular and therefore the applied voltage is significantly reduced).

The ease with which equation (\ref{eqn:geometric_condition}) can be satisfied suggests that not only will we have $v_{rh}<0$ in a typical accretion disk, but the accretion flow may also reach a steady state in which the main, turbulent body of the disk drags the nonturbulent surface layer inward at the same speed as itself; i.e., $\left| v_{rh} \right| \sim \left< v_{r} \right>$.  In fact, if we start with equation (\ref{eqn:vrh}) and go through the same analysis as above but require $v_{rh} \leq - \left< v_{r} \right>$ rather than $v_{rh}<0$, then instead of equation (\ref{eqn:geometric_condition}) we obtain
\begin{equation}
-\frac{B_{\phi h} B_{z h}}{\mathcal{B}_{zh}^{2}} \gtrsim \frac{H_{B}}{H} \frac{\left< v_{r} \right>}{H \Omega_{K}} \left[ 1 + \frac{3h}{r} \frac{v_{zh}}{\left< v_{r} \right>} \right] ,
\label{eqn:drag_same_speed}
\end{equation}
where the right hand side is typically $\sim \alpha H_{B}/r$ for a dynamically weak field.  Like equation (\ref{eqn:geometric_condition}), this condition is modest assuming the disk is thin, and thus we may expect that advection of magnetic fields proceeds at the same speed as turbulent accretion in the main body of the disk.  In fact, equation (\ref{eqn:drag_same_speed}) suggests that advection of magnetic fields in the surface layer could proceed {\it faster} than the disk accretion speed, but as we discuss in Appendix \ref{appendix:geometry}, this is unlikely to be sustainable.

The only qualification to what we have said so far concerns the sign of $B_{\phi h} B_{zh}$.  In particular, equation (\ref{eqn:bcondition}) shows that $B_{\phi h} B_{zh} \lesssim 0$ is required in order for the field to advect inward; the exact condition is $\partial \left( B_{\phi} B_{z} \right)_{h} / \partial z < 0$, which states that the large-scale magnetic field must remove angular momentum from the nonturbulent surface layer.  This is a strict requirement for disks in which $v_{zh} \geq 0$.  We expect that $\partial \left( B_{\phi} B_{z} \right)_{h} / \partial z < 0$ will be satisfied in many parts of an accretion disk, but not necessarily all.  It will be satisfied in regions where the magnetic field has a dipole-type symmetry (with $B_r$ and $B_\phi$ odd functions of $z$ and $B_z$ an even function), which is often assumed for the large-scale magnetic field advected inward in an accretion disk (see Figure \ref{fig:hourglass}).  However, in a region of the disk with a quadrupole-type field symmetry (with $B_r$ and $B_\phi$ even functions of $z$ and $B_z$ an odd function), as may occur when the large-scale field extending out of the disk is generated primarily by magnetorotational turbulence \citep[e.g.,][]{Brandenburg1995}, some regions will likely have $\partial \left( B_{\phi} B_{z} \right)_{h} / \partial z > 0$.  In these regions, angular momentum will not be removed vertically from the surface of the disk, and inward radial advection of the magnetic field may be difficult to sustain.  Correspondingly, we find that one of the simulations described in detail in \citet{MillerStone2000} has $\partial \left( B_{\phi} B_{z} \right)_{h} / \partial z < 0$, but the other does not, so that the material (and magnetic field) in the nonturbulent region may not advect inward in this second case.

On the other hand, if there is a weak vertical seed field in the outer part of the disk, shear and MRI turbulence will create azimuthal field from it, and the condition $\partial \left( B_{\phi} B_{z} \right)_{h} / \partial z < 0$ should be satisfied automatically, while the condition in equation (\ref{eqn:geometric_condition}) will be unchanged.  We therefore view this as the most favorable way to induce magnetic field advection in a disk.  The large-scale magnetic field introduced into the disk at large distance may be supposed to come from the interstellar medium (in the case of supermassive black holes) or from a companion star (in the case of X-ray binaries and other stellar-mass systems).

To conclude this section, we consider the second regime for magnetic field advection alluded to in our earlier discussion of equation (\ref{eqn:condition}), in which the geometry is such that $\left( B_{rh} / B_{zh} \right) v_{zh} < 0$, and inward radial advection requires $v_{rh} < \left( B_{rh} / B_{zh} \right) v_{zh}$ in order to overcome vertical advection and produce a concentration of magnetic flux in the inner region of the disk.  If the disk is thin, with $3h/r \ll \left| B_{rh} / B_{zh} \right|$, a sufficient condition for this to occur can be obtained by replacing $h/r$ with $- B_{rh} / 3 B_{zh}$ in equation (\ref{eqn:bcondition}) and propagating this change through subsequent expressions.  In particular, equation (\ref{eqn:geometric_condition}) becomes
\begin{equation}
-\frac{B_{\phi h} B_{z h}}{\mathcal{B}_{zh}^{2}} \gtrsim \frac{H_{B}}{H} \left( - \frac{B_{rh}}{B_{zh}} \right) \frac{ v_{zh} }{H \Omega_{K}} ,
\end{equation}
where we have $\left( - B_{rh}/B_{zh} \right) v_{zh} > 0$ by definition so that the right hand side of the equation is, in general, positive.  This condition can be satisfied for $\left| B_{rh} / B_{zh} \right| \sim 1$ provided that $v_{zh}$ is not too close to the sound speed.

\subsection{Conditions on the Magnetic Field Strength and Ionization Fraction} \label{sec:strength}

From the analysis of the previous subsection, an arbitrarily weak seed field threading an accretion disk should be able to advect inward along the disk surface.  But what really happens for arbitrarily weak fields?  Are there field strengths below which some of our underlying assumptions in this paper break down?

An important assumption in this paper is that the region above the disk is nonturbulent and, therefore, highly conducting; it is this region in which the magnetic field can advect inward.  Clearly, a nonturbulent region is likely to exist {\it somewhere} above an accretion disk, but the question is whether it occurs at a low enough height to be treated as the ``surface layer'' of the disk, as we do in this paper. Equation (\ref{eqn:beta_condition}) and the usual assumption that the mass density decreases with height much more rapidly than the magnetic energy density suggests that the nonturbulent region should occur within a few scale heights, even for a very weak seed field on the equatorial plane.  However, if the magnetic energy density begins to drop off rapidly with height, the turbulence may not be suppressed until a very large distance above the disk.  This may be what happens in the radiation-dominated simulations of \citet{Turner2004}, where the magnetic energy density begins to fall off at $z \gtrsim 3H$, and there is no clear evidence for a nonturbulent region anywhere within the simulation domain (which extends out to $z \sim 8H$).

It is difficult to predict when this type of behavior will occur, but when
it does, our assumption that the nonturbulent region occurs ``within the disk'' may break down.  In particular, orbits may not be circular, so that equation (\ref{eqn:vrh}) is no longer strictly valid.  Considering radial force balance and using equation (\ref{eqn:beta_condition}), we find that if the disk is sufficiently thin, magnetic forces are unlikely to be strong enough to disrupt circular orbits at the base of the nonturbulent region (although they may certainly do so higher up in the corona); the condition for magnetic forces to be negligible at $z \approx h$ is $\mathcal{B}_{zh}^{2} / B_{h}^{2} \gg \left( H/r \right)^{2}$ and $\mathcal{B}_{zh}^{2} / \left| B_{rh} B_{zh} \right| \gg H/r$, both of which are easily satisfied by the \citet{MillerStone2000} simulations (here $\mathcal{B}_{zh}^{2} \approx B_{zh}^{2}$ was defined in \S \ref{sec:geometry}, and in the second expression we have assumed that the scale height of $\left| B_{rh} B_{zh} \right|$ can be approximated as $\sim H$).  Radiation pressure is therefore the only realistic concern; in order for orbits to remain circular, we require $\rho_{h} v_{K}^{2}$ to dominate over the radiation pressure.  Assuming the temperature at $z \approx h$ is given by the effective surface temperature of a \citet{ShakuraSunyaev1973} disk, we can use equation (\ref{eqn:beta_condition}) to derive
\begin{equation}
\mathcal{B}_{zh} \gg 10^{-2} \; m_{8}^{-1/2} \dot{m}^{1/2} \hat{r}_{3}^{-3/2} f_{*}^{1/2} \left( 10^{2} H/r \right) \; {\rm G}
\label{eqn:circular_orbits}
\end{equation}
as the condition for circular orbits, where $\hat{r}_{3} \equiv \hat{r} / 10^{3}$ and $m_{8} \equiv m / 10^{8}$ (i.e, we have scaled the fiducial value to that which would occur at a distance of $10^{3}$ Schwarzschild radii from a supermassive black hole of mass $10^{8} M_{\sun}$).

\begin{figure}
\epsscale{1.1} \plotone{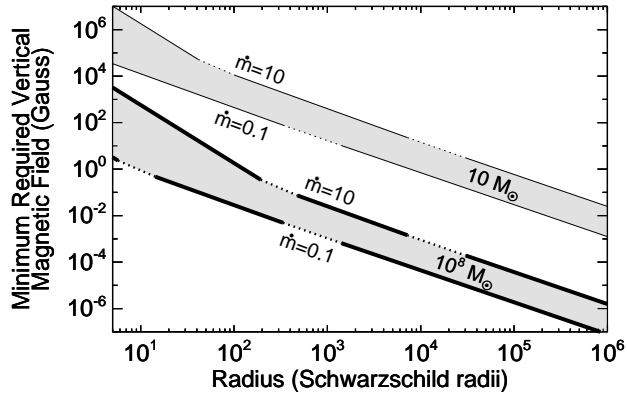} \epsscale{1.0}
\caption{Minimum values of the vertical magnetic field $\mathcal{B}_{zh}$ that are required in order for the large-scale field to overcome radiation pressure and be advected inward in the nonturbulent surface layer of an accretion disk.  Constraints are shown for black hole masses of $10 M_{\sun}$ and $10^{8} M_{\sun}$ and dimensionless accretion rates between $\dot{m}=0.1$ and $\dot{m}=10$ (note that the Eddington accretion rate corresponds to $\dot{m} = \epsilon^{-1}$, where $\epsilon$ is the radiative efficiency of the disk and $\epsilon \approx 0.06$ for a simplified treatment of accretion onto a Schwarzschild black hole).  We assume a \citet{ShakuraSunyaev1973} disk with $f_{*}=1$ at the inner boundary and $\alpha=10^{-4}$ as a worst-case lower limit for the turbulent stress.  Solid lines show regions in which the \citet{ShakuraSunyaev1973} assumptions about pressure and opacity sources in the disk are good to at least 50\%; dotted lines are used to connect through transition regions in which the \citet{ShakuraSunyaev1973} solution breaks down.\\}
\label{fig:strength}
\end{figure}

\begin{figure}
\epsscale{1.1} \plotone{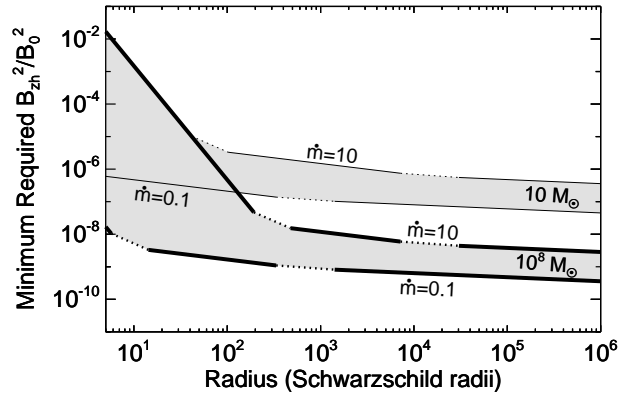} \epsscale{1.0}
\caption{Same as Figure \ref{fig:strength}, but for the minimum values of the ratio $\mathcal{B}_{zh}^{2}/B_{0}^{2}$ (between the energy density of vertical magnetic fields at the disk surface and turbulent fields on the equatorial plane) that are required in order for advection of the large-scale magnetic field to overcome radiation pressure.\\}
\label{fig:geometry}
\end{figure}

Limits on the magnetic field strength resulting from equation (\ref{eqn:circular_orbits}) are plotted in Figure \ref{fig:strength} for typical \citet{ShakuraSunyaev1973} accretion disks with $\alpha \approx 10^{-4}$, which we take to be a worst-case lower limit for the turbulent stress (note in any case that the dependence on $\alpha$ is very weak).  In fact, $\alpha$ is not independent of the field strength; using the relation $\alpha \sim 0.5 B_{0}^{2} / 8 \pi p_{0}$ between turbulent magnetic stress and turbulent magnetic energy density $B_{0}^{2}/8 \pi$ in the main body of the disk \citetext{which is a robust result of MRI simulations; e.g., \citealt{Hawley1995}; \citealt{Sano2004}; \citealt*{Blackman2006}}, we can rewrite equation (\ref{eqn:circular_orbits}) as a constraint on the field geometry:
\begin{equation}
\frac{\mathcal{B}_{zh}^{2}}{B_{0}^{2}} \gg 4 \times 10^{-6} \left( \frac{\alpha \rho_{0}}{{\rm g \; cm^{-3}}} \right)^{-1} m^{-1} \dot{m} \hat{r}^{-2} f_{*} .
\label{eqn:circular_orbits_geometric}
\end{equation}
Limits on this ratio for typical \citet{ShakuraSunyaev1973} accretion disks are plotted in Figure \ref{fig:geometry}, again assuming $\alpha \approx 10^{-4}$.  In the radiation-dominated region of the disk, $\rho_{0} \propto \alpha^{-1}$, so the right hand side of equation (\ref{eqn:circular_orbits_geometric}) has no dependence on the magnetic field strength, whereas in the gas pressure-dominated region, the dependence occurs through $\alpha^{-3/10}$, which we have fixed to the assumed worst-case value.  Thus, although equation (\ref{eqn:circular_orbits_geometric}) technically represents a joint limit on the equatorial plane field strength and the vertical magnetic geometry, the condition on the geometry is more important.

In a conservative analysis, equation (\ref{eqn:circular_orbits}) can be compared to the large-scale magnetic field that might be supplied to the outer disk (from, say, the interstellar medium or a companion star) in order to determine whether or not orbits are circular in this region.  However, equation (\ref{eqn:circular_orbits_geometric}) may be a more appropriate expression if we allow for the possibility that buoyant rising of magnetic fields from the turbulent disk into the nonturbulent corona can affect the magnitude of $\mathcal{B}_{zh}$; for example, the simulations of \citet{MillerStone2000} have $\mathcal{B}_{zh}^{2} / B_{0}^{2} \gtrsim 0.02$ (even without any net imposed vertical field) and thus should easily satisfy equation (\ref{eqn:circular_orbits_geometric}).

As mentioned in \S \ref{sec:coupling}, even when orbits are not circular, equation (\ref{eqn:vrh}) and the subsequent analysis may still be valid to an order of magnitude, and so we do not view the constraints on the magnetic field discussed here as fundamental lower limits.  On the other hand, one might reasonably expect that the effect of radiation pressure is to drive material radially outward from the hot inner disk and thereby prevent the nonturbulent surface layer and its associated magnetic field from accreting; in that case, circular orbits can indeed be viewed as a strict requirement.  Conversely, other mechanisms that might produce noncircular orbits, such as advection-dominated flows at either high or low accretion rates \citep{Abramowicz1988,NarayanYi1994,NarayanYi1995}, tend to enhance the inward radial velocity and therefore do not prevent inward advection of magnetic fields from occurring.  In these cases, equation (\ref{eqn:geometric_condition}) should have $v_{K}$ replaced by the actual azimuthal velocity $v_{\phi}$, which makes the equation more difficult to satisfy but does not change our overall conclusions.

Even if orbits in the nonturbulent region are circular, other constraints may come about that could affect the ability of this layer to advect magnetic fields inward.  In particular, our discussion so far has implicitly assumed that the disk is fully ionized.  We do not discuss partially ionized disks in depth in this paper, but we point out two important issues.  First, the diffusivity associated with electrons scattering off of neutrals (as well as other non-ideal MHD effects) may become more important than the electron-ion Spitzer diffusivity, and second, there may not be enough free electrons in the nonturbulent surface layer of the disk to support the advected magnetic field.

We consider non-ideal MHD effects first.  Following \citet{PandeyWardle2007}, a more general version of equation (\ref{eqn:induction_equation}) which includes the relevant non-ideal MHD terms is
\begin{equation}
\frac{\partial{\bf B}}{\partial t} = {\bf \nabla} \times \left[ {\bf v}^{\prime} {\bf \times B} - \frac{4 \pi \eta}{c} {\bf J} + \frac{\rho_{n}^{2} \left( {\bf J \times B} \right) {\bf \times B}}{\rho^{2} \rho_{i} \nu_{in} c} \right] ,
\label{eqn:full_induction_equation}
\end{equation}
where ${\bf v}^{\prime} = {\bf v} - {\bf J} / n_{e} e$ is the plasma velocity modified by the Hall drift, ${\bf J} \equiv \left( c / 4 \pi \right) {\bf \nabla \times B}$ is the current density, $n_{e}$ is the electron number density, $e$ is the proton charge, $\rho_{n}$ and $\rho_{i}$ are the neutral and ion mass densities, and $\nu_{in}$ is the ion-neutral collision frequency.  The last two terms in equation (\ref{eqn:full_induction_equation}) represent the effects of Ohmic diffusion (i.e., the scattering of electrons off of ions and neutrals) and ambipolar diffusion, respectively.

We use equations in \citet*{Draine1983} and \citet{BalbusTerquem2001} to estimate the importance of the non-ideal MHD effects.  For simplicity, we assume {\it a priori} that the plasma is weakly ionized and that the electron and ion number densities are the same.  We take ``worst-case'' approximations of quantities involving the magnetic field (for example, $\left| {\bf \nabla \times B} \right|_{h} \sim B_{h} / H$), assume that the plasma temperature is given by the effective surface temperature of a \citet{ShakuraSunyaev1973} disk, and make use of equation (\ref{eqn:beta_condition}) to substitute for the density when needed.  Fiducial values are given for a distance of $10^{3}$ Schwarzschild radii from a $10^{8} M_{\sun}$ black hole, where the surface temperature is of order 2,000 K and thus the disk should be weakly ionized.  For diffusive processes, we can calculate an effective magnetic Reynolds number $Re_m$ in the nonturbulent surface layer of the disk (see \S \ref{sec:advection}).  When $Re_m \gg 1$, advection of the magnetic field will dominate.  For Ohmic diffusion, we find that this condition can be approximately written as
\begin{equation}
\frac{n_{i}}{n_{n}} \gg 5 \times 10^{-15} \; m_{8}^{-9/8} \hat{r}_{3}^{-7/8} D_{Ohm} ,
\label{eqn:ohmic}
\end{equation}
where $n_{i}/n_{n}$ is the ionization fraction at the base of the nonturbulent region (i.e., the ratio of ion to neutral number densities) and we have defined a parameter $D_{Ohm} \equiv \left( 10 \alpha \right)^{-1} \dot{m}^{1/8} f_{*}^{9/8} \left( 10^{2} H/r \right)^{-3} \mathcal{U}_{s}^{-1}$ which contains terms of lesser importance.  Here, $\mathcal{U}_{s} \equiv \left| v_{rh} \right| / \left< v_{r} \right>$ is likely to be $\approx 1$ as per our discussion in \S \ref{sec:geometry}.   For ambipolar diffusion, the condition becomes
\begin{equation}
\frac{n_{i}}{n_{n}} \left( \frac{\mathcal{B}_{zh}}{\rm G} \right)^{2} \gg 10^{-3} \; m_{8}^{-1} \hat{r}_{3}^{-5/2} D_{amb} ,
\label{eqn:ambipolar}
\end{equation}
where $D_{amb} \equiv \left( 10 \alpha \right)^{-1} f_{*} \left( 10^{2} H/r \right) \mu^{\prime} \left| B_{h}/5 \mathcal{B}_{zh} \right|^{2} \mathcal{U}_{s}^{-1}$ and $\mu^{\prime} \equiv \mu \left( 1 + m_{n}/m_{i} \right)$; here, $\mu$ is the mean mass per particle expressed in units of the proton mass and $m_{n}/m_{i}$ is the ratio of neutral to ion masses.

We therefore see that ambipolar diffusion imposes a joint constraint on the ionization fraction and magnetic field strength that is generally much more important than the limit imposed by Ohmic diffusion; this is a direct result of the low densities expected in the nonturbulent surface layer that we consider in this paper.  For a partially ionized disk that is able to overcome Ohmic diffusion and satisfy equation (\ref{eqn:ohmic}), magnetic fields can advect inward, but equation (\ref{eqn:ambipolar}) shows that stronger initial seed fields than those indicated in equation (\ref{eqn:circular_orbits}) and Figure \ref{fig:strength} (for the constraint imposed by circular orbits) are generally required.  However, in these applications it is important to keep in mind that the ionization fraction that appears in the above equations is calculated at the surface of the disk (as opposed to the main disk body), where interstellar cosmic rays may inflate the ionization rate above the values that one would usually expect \citep{Gammie1996}.

From equation (\ref{eqn:full_induction_equation}), we see that the Hall effect, which occurs due to the drift of the magnetic field with respect to the ions as it is carried along by the electrons, may also be important, not only for partially ionized disks but also for fully ionized ones.  However, the Hall effect does not cause diffusion of the magnetic field and therefore does not necessarily oppose inward advection.  In fact, if the radial component of the current density is positive at the surface of the disk, Hall drift enhances inward advection rather than opposes it; this occurs when $\partial B_{\phi h} / \partial z < 0$, a condition which is already met by the dipole-type fields discussed in \S \ref{sec:geometry}.  In geometries where the Hall effect opposes inward advection, however, the condition for it to be negligible is
\begin{equation}
\frac{n_{i}}{n_{n}} \left( \frac{\mathcal{B}_{zh}}{\rm G} \right) \gg 2 \times 10^{-8} \; m_{8}^{-1} \hat{r}_{3}^{-3/2} D_{Hall} ,
\label{eqn:hall}
\end{equation}
where $D_{Hall} \equiv \left( 10 \alpha \right)^{-1} f_{*} \left( 10^{2} H/r \right)^{-1} \mu \left| B_{h}/5 \mathcal{B}_{zh} \right| \mathcal{U}_{s}^{-1}$.  This equation also applies for a fully ionized disk if we replace $n_{i}/n_{n}$ by $\approx 1/2$.  Note that the Hall effect is generally more important in a fully ionized accretion disk than the electron-ion Spitzer diffusivity discussed in \S \ref{sec:advection}, and in some cases (especially for stellar mass objects), it may even impose a stronger constraint on the magnetic field strength than that imposed by equation (\ref{eqn:circular_orbits}) for circular orbits.

We finally consider the possibility that in the diffuse surface layer of a weakly ionized disk, there may not be enough free electrons available to carry the current that is necessary for supporting the large-scale magnetic field.  Making similar assumptions as above and also assuming that the velocity of electrons is limited to their thermal speed, we find that
\begin{equation}
\frac{n_{i}}{n_{n}} \left( \frac{\mathcal{B}_{zh}}{\rm G} \right) \gtrsim 9 \times 10^{-12} \left[ \frac{\mu \left( 10^{2} H/r \right) \left| B_{h}/5 \mathcal{B}_{zh} \right|}{m_{8}^{7/8} \dot{m}^{1/8} \hat{r}_{3}^{13/8} f_{*}^{1/8}} \right] 
\end{equation}
is required to produce enough current to support the magnetic field, which is generally a weaker constraint than that imposed by ambipolar diffusion.  (Note that stronger magnetic fields are actually {\it easier} for the nonturbulent surface layer to support; this is because strong fields suppress turbulence at a lower height above the disk, where the electron density is larger.)

\subsection{The Outcome of Advection} \label{sec:outcome}

In summary, we find that weak, large-scale magnetic fields can be advected inward in the surface layer of an accretion disk.  The most important condition necessary for this is $\partial \left( B_{\phi} B_{z} \right)_{h} / \partial z < 0$ (i.e., the vertical magnetic stress must extract angular momentum from the disk surface).  This may occur most easily if a weak, dipole-type seed field is supplied at the outer regions of the disk.  This field threads the disk and acts as a catalyst for production of the toroidal fields that in turn provide the required geometry for further inward advection.

Although the constraints on the field strength discussed in \S \ref{sec:strength} are modest, we can nonetheless imagine that a weak, dipole-type magnetic field (relatively close to the lower limit) is supplied in the outer region of the disk and ask the following question:  How easy will it be for the field to remain at the required level as it is advected inward and compressed?  From equation (\ref{eqn:circular_orbits}), we find that roughly, the vertical field must increase faster than $B_{z} \propto r^{-3/2}$ in the gas pressure-dominated region of a \citet{ShakuraSunyaev1973} disk in order to satisfy the requirement for advection to continue; this limit becomes $B_{z} \propto r^{-5/2}$ in the inner, radiation-dominated region (where $H$ is approximately constant).  This growth seems difficult to sustain, which suggests the interesting possibility that an advected magnetic field may temporarily ``stall'' at some large radius when it becomes too weak to advect inward, and advection will only continue once enough magnetic flux has built up at this location.  On the other hand, as we have seen in discussing equation (\ref{eqn:circular_orbits_geometric}), even if the advected field does not increase in strength fast enough on its own, it may be reasonable to assume that the local dynamo can produce strong enough surface fields (through magnetic buoyancy) to meet these conditions, and thus advection will continue in either case.

As discussed in \S \ref{sec:intro}, a sufficiently strong magnetic field threading the disk can lead to inward radial advection of the field driven by the {\it main} body of the disk too, due to extraction of angular momentum from the main body of the disk to a wind or jet \citep{Lovelace1994}.  However, the field strengths required for this process are much larger; the field must drive accretion in the dense body of the disk (rather than in a low-density surface layer).  In fact, advection of magnetic fields in the surface layer is always more efficient, provided that the disk is thin and the vertical field is subequipartition.

Although advection in the main body of the disk must take place fast enough to overcome turbulent diffusion, advection in the nonturbulent region is limited by the much smaller speeds associated with the microscopic diffusivity.  Magnetic fields can advect inward in the nonturbulent surface layer even when $\left| v_{rh} \right| \ll \left< v_{r} \right>$, and the advection can therefore take place on timescales longer than a viscous timescale.  This will lead to a gradual buildup of field in the inner disk.  In the case of angular momentum extraction from the main body of the disk, advection of magnetic fields always occurs at {\it least} as fast as the viscous timescale, and sustained advection requires that the magnetic field be strong enough to completely overwhelm turbulent diffusion; otherwise, a steady state will be reached in which only a small concentration of magnetic field develops in the inner disk.  Thus, we may expect a disk to experience gradual advection of large-scale magnetic fields on long timescales in the surface layer, with occasional brief bursts of fast or implosive accretion \citep{Lovelace1994} associated with the presence of a strong magnetic field in a local region of the disk that extracts angular momentum from the main disk body.

It is interesting to note that advection of magnetic fields in the surface layer is more efficient for thin disks than thick ones, as can be seen, for example, from equation (\ref{eqn:bcondition}).  This is because the surface layer is tightly coupled to the main body of the disk rather than being part of a detached corona.  On the other hand, turbulent advection of magnetic fields in the main body of the disk is more efficient if the disk is thick (since $Re_m \approx H/r$, as discussed in \S \ref{sec:advection}).  We therefore conclude that magnetic field advection is possible for many disks but most efficient for thin ones.

\section{Conclusions} \label{sec:conclusions}

This paper reanalyzes the advection of a large-scale, weak magnetic field in an accretion disk.  We consider the vertical structure of the disk, which strongly influences the vertical profile of the conductivity, as pointed out by \citet{BKL2007}.  In the thin, diffuse surface layers of the disk, the magnetic energy density is large enough compared to the thermal energy density that magnetorotational turbulence is suppressed.  As a consequence, magnetic field lines threading the surface layer can be advected inward with the main body of the disk, without being opposed by turbulent diffusion.

No special conditions are required for the field to be advected inward
except that it meet the rather modest constraints in \S \ref{sec:conditions}.  The required field strengths are relatively weak, and the primary constraint on the field geometry is simply that it must help the nonturbulent surface layer accrete inward (i.e., the vertical magnetic stress must extract angular momentum from this layer, in virtually any amount).  This can be accomplished either via coupling between the main, turbulent body of the disk and the surface, or via a wind or jet.  The simplest way for this condition to be met is if the accretion flow is provided with a weak, large-scale vertical seed field threading the outer region of the disk (which could come from the interstellar medium or a companion star), although in some cases, the proper geometry may be attained entirely as a result of fields produced via the local magnetorotational dynamo.  Once a weak, large-scale field with the proper geometry is in place, the field will be advected inward along the disk's surface layer and strengthened as it is compressed along with the accretion flow.

The presence of a large-scale magnetic field anchored in the surface layer will drive strong magnetorotational turbulence in the main body of the disk, which can consequently produce values of the $\alpha$ parameter (i.e., the turbulent stress) large enough to match observational constraints.  We find that typical vertical fields on the order of a few percent of equipartition are required for this to occur.  Because the field is advected inward and anchored in the surface layer, there is no need to worry about maintaining the required vertical fields via internal magnetorotational fluctuations \citep[as suggested by][]{Pessah2007}.  We propose that long-term changes in $\alpha$ (in response to the history of magnetic field advection) should be explored as a possible source of the long-term evolution in the variability patterns seen in the light curves of X-ray binaries such as GRS~1915+105 \citep[see, e.g.,][]{Tagger2004}.

The mechanisms we discuss in this paper are relevant to many different kinds of accreting objects.  In the outer part of the disk (far away from the central star or black hole), our work should be applicable provided only that the disk is sufficiently ionized (see \S \ref{sec:strength}).  Closer to the inner part of the disk, our work is most obviously applicable to black holes, where the large-scale magnetic field arises entirely within the accreting plasma.  However, it may also be relevant to the case where a large-scale field from a magnetized central star penetrates the disk; previous suggestions that outward radial diffusion of the field may be an important process in such systems \citep*[e.g.,][]{Lovelace1995,BardouHeyvaerts1996,Agapitou2000,Uzdensky2002} should be examined in light of our work.  Finally, the applicability of our work to radiation-dominated regions of the disk may require further analysis, because it is unclear whether the vertical structure of these regions can support the inward advection of magnetic fields (although we argue here that it can).  Future numerical simulations may eventually be able to address this point.

Also, we have assumed axisymmetric disks in this paper, but the equations we have derived apply more generally if they are averaged over azimuth (with the introduction of appropriate correction factors).  Thus, a magnetic field with an {\it azimuthally-averaged} strength and {\it azimuthally-averaged} geometry that roughly meets the criteria in \S \ref{sec:conditions} is likely to be able to advect inward in an accretion disk.  Nonaxisymmetric advection of a large-scale magnetic field \citep[e.g., as envisioned by][]{Spruit2005} is clearly consistent with the mechanism we have proposed in this paper, but we have also shown that advection is equally plausible in an axisymmetric disk, without having to assume any special conditions.  More numerical simulations that investigate disks in which a large-scale magnetic field is supplied at the outer boundary are clearly needed.

Finally, our work has important implications for models of jet formation which require strong, large-scale magnetic fields to exist over a region of the inner accretion disk.  As suggested more than 30 years ago, these magnetic fields can arise in the inner disk via a very simple process: advection of a weak field from outside.  This opens up the possibility for magnetically-dominated outflows (i.e., Poynting jets) to exist in the inner regions of disks around a wide variety of accreting objects.  In addition, the radial stretching of field lines produced by advection may allow winds accelerated by the magnetocentrifugal effect \citep{BlandfordPayne1982} to exist over a wide range of radii.  This is in contrast to the results of numerical simulations in which no magnetic fields are supplied at the outer boundary, and jets only form in an extremely narrow inner region where the energetics may be dominated by relativistic effects that require the presence of a rotating black hole.

\section*{Acknowledgements}

We thank G.~S. Bisnovatyi-Kogan, M.~M. Romanova, and I.~G. Igumenshchev for valuable discussions.  D.~M.~R. is supported by an NSF Astronomy and Astrophysics Postdoctoral Fellowship under award AST-0602259.  The work of R.~L. was supported in part by NASA grants NAG5-13220 and NAG5-13060 and by NSF grant AST-0507760.

\begin{appendix}

\section{Physical Conditions at the Base of the Nonturbulent Region Above an Accretion Disk} \label{appendix:density}

In this section we derive equation (\ref{eqn:beta_condition}), which defines the base of the nonturbulent surface layer of the accretion disk and is used throughout the main body of the paper.

We are interested in conditions at $z=h$, the height in the disk where the magnetorotational instability (MRI) is first suppressed.  The fundamental condition for suppression of the MRI is that the Alfv{\'e}n speed must be large enough so that fluid elements linked by the magnetic field will be drawn back together faster than orbital shear drives them apart; in a WKB (small wavelength) analysis, MRI modes with wavenumber ${\bf k}$ are found to be suppressed when ${\bf k \cdot v_{A}} \gtrsim \Omega_{K}$, where $v_{A}$ is the local Alfv{\'e}n speed \citep[e.g.,][]{BalbusHawley1998} and we have assumed circular orbits (relaxing this assumption would simply replace $\Omega_{K}$ with the actual local angular velocity of the disk).  If we take the WKB approximation to its limit and make the usual assumption that the MRI will be completely suppressed when no unstable wavelength fits within the disk \citep[e.g.,][]{BalbusHawley1991} and further consider that the largest wavelengths that fit within the disk are of order $\sim \left( {\rm a \; few} \right) \times H$ in the vertical direction and $\sim \left( {\rm a \; few} \right) \times r$ in the azimuthal direction, then applying this condition to the material at $z=h$ shows that MRI turbulence will be suppressed when
\begin{equation}
\left( v_{Az} \right)_{h} + \frac{H}{r} \left( v_{A \phi} \right)_{h} \gtrsim H \Omega_{K} ,
\label{eqn:valf}
\end{equation}
where $\left( v_{Az} \right)_{h} \equiv B_{zh} / \sqrt{4 \pi \rho_{h}}$ and $\left( v_{A \phi} \right)_{h}$ is defined equivalently.  Rearranging this equation gives, approximately,
\begin{equation}
\rho_{h} \lesssim \frac{\mathcal{B}_{zh}^{2}}{8 \pi H^{2} \Omega_{K}^{2}} ,
\label{eqn:beta_condition_appendix_first}
\end{equation}
where the equality holds at $z \approx h$ (the height at which turbulence is first suppressed), but the equation also applies more generally if $h$ is redefined to be any height within the nonturbulent region.  Here, we have defined a modified vertical magnetic field strength $\mathcal{B}_{zh} \equiv \left| B_{zh} \right| + \left( H/r \right) \left| B_{\phi h} \right|$.  In the main body of the paper, we generally assume $\mathcal{B}_{zh} \approx \left| B_{zh} \right|$, but the full expression should be used when the toroidal field is extremely strong \citetext{$\mathcal{B}_{zh} \approx \left| B_{zh} \right|$ appears to be a good approximation for the magnetic fields seen in the simulations of \citealt{MillerStone2000}, however}.

If we define $p_{0} \equiv \rho_{0} \left( H \Omega_{K} \right)^{2}$, equation (\ref{eqn:beta_condition_appendix_first}) becomes
\begin{equation}
\frac{\rho_{h}}{\rho_{0}} \lesssim \frac{\mathcal{B}_{zh}^{2}}{8 \pi p_{0}} ,
\label{eqn:beta_condition_appendix}
\end{equation}
which is the equivalent of equation (\ref{eqn:beta_condition}).  If the large-scale magnetic field is weak enough so as to not significantly affect the dynamics in the main body of the disk (in particular, if thermal pressure supports the disk vertically against gravity), then $p_{0}$ which appears in equation (\ref{eqn:beta_condition_appendix}) should be interpreted as the thermal pressure.  However, the assumption of weak fields is not required, and equation (\ref{eqn:geometric_condition}), which is derived from equation (\ref{eqn:beta_condition_appendix}), applies when the field is strong as well as when it is weak.  Only in the parts of the main body of the paper where we combine equation (\ref{eqn:beta_condition}) with the \citet{ShakuraSunyaev1973} solution are we assuming that the large-scale magnetic field is dynamically weak.

Equation (\ref{eqn:beta_condition_appendix}) is not the same as assuming that the magnetic and thermal energy densities are comparable at $z \approx h$, which is sometimes quoted as the condition for suppression of the MRI.  We note, however, that the condition on the Alfv{\'e}n speed is more fundamental, and the condition on the magnetic energy density is merely derived from it under a specific set of circumstances in the main body of the disk.  In fact, analytical studies of both stratified and unstratified disks generally suggest that the condition ${\bf k \cdot v_{A}} \gtrsim \Omega_{K}$  is most important for MRI suppression, regardless of the overall magnetic energy density \citep{BlaesBalbus1994,GammieBalbus1994,KimOstriker2000}.  Thus, we believe that equation (\ref{eqn:beta_condition_appendix}) is correct.

In fact, what is most clear from the above MRI studies is that the stability criteria in each coordinate direction are roughly independent; for example, a strong toroidal field does not significantly affect the most unstable vertical wavelengths.  Thus, if we {\it were} to suppose that the MRI is suppressed when the local magnetic and thermal energy densities are equal, it would be reasonable to assume that the appropriate condition is $B_{zh}^{2} \gtrsim 8 \pi p_{h}$ (i.e., that it involves the vertical rather than the total magnetic energy density), in which case we would derive $\left( p_{h} / p_{0} \right) \lesssim B_{zh}^{2} / 8 \pi p_{0}$ rather than equation (\ref{eqn:beta_condition_appendix}) as the condition for MRI suppression; here, the subscripts have their usual meanings.  We therefore see that the two possible conditions are essentially the same for a disk that is dominated by gas pressure up to $z \approx h$; one would simply need to modify equation (\ref{eqn:beta_condition_appendix}) by introducing a ratio of temperatures $T_{0} / T_{h}$ on the right hand side, which is generally no more than a factor of a few.  Only if the disk is radiation-dominated at $z \approx h$ is there a significant difference between the two conditions for suppression of MRI turbulence.  In the case of a radiation-dominated disk, our use of equation (\ref{eqn:beta_condition_appendix}) means that we are assuming the MRI may be suppressed in the surface layer of the disk even when the magnetic pressure is weaker than the radiation pressure there.

\section{Magnetic Field Advection in Quasi-Stationary Accretion Disks} \label{appendix:geometry}

Our work in this paper is concerned with accretion disks that are fundamentally time-dependent.  In particular, we are studying situations in which a large-scale magnetic field is being dragged inward and causing the magnetic flux in the inner region of the disk to change in accordance with equation (\ref{eqn:magnetic_flux}).  Such disks are complicated to study analytically, and for this reason we used order of magnitude approximations to reach some of our conclusions in \S \ref{sec:geometry} (in particular, to estimate $H_{B}$, the scale height of the vertical magnetic stress).

In this section, we will give arguments for these approximations by considering the simplified case of a ``quasi-stationary'' disk.  By this we mean a disk that is allowed to vary on the long (viscous) timescales at which magnetic field advection takes place but which is assumed to quickly adjust its structure on shorter timescales (in response to the changing magnetic flux) so that in most respects it can be treated as stationary.

We begin by evaluating the $\phi$ component of the induction equation (\ref{eqn:induction_equation}) at $z=h$ (where the magnetic diffusivity is negligible) in an axisymmetric disk, which gives
\begin{equation}
\frac{\partial B_{\phi h}}{\partial t} = \frac{\partial}{\partial z} \left( v_{\phi} B_{z} - v_{z} B_{\phi} \right)_{h} - \frac{\partial}{\partial r} \left( v_{r} B_{\phi} - v_{\phi} B_{r} \right)_{h} .
\label{eqn:phi_induction_initial}
\end{equation}
If we assume locally circular orbits in Newtonian gravity and evaluate spatial derivatives of the orbital velocity $v_{\phi}$ to first order in $h/r$, we can combine equation (\ref{eqn:phi_induction_initial}) with the condition ${\bf \nabla \cdot B} = 0$ to derive
\begin{equation}
\frac{\partial B_{\phi h}}{\partial t} \approx - \frac{3}{2} \Omega_{K} \left( B_{rh} + \frac{h}{r} B_{zh} \right) - \frac{\partial}{\partial z} \left( v_{z} B_{\phi} \right)_{h} - \frac{\partial}{\partial r} \left( v_{r} B_{\phi} \right)_{h} .
\label{eqn:phi_induction}
\end{equation}
The first term in this equation represents the production of azimuthal field via Keplerian shear and is generally the most important effect; it leads to the creation of $B_{\phi}$ on a characteristic timescale of $\sim \Omega_{K}^{-1}$ (i.e., the orbital timescale).  Therefore, in accordance with our quasi-stationary approximation, we can assume that the disk quickly adjusts its value of $B_{\phi h}$ in response to the magnetic field advection so that $\partial B_{\phi h} / \partial t$ is negligible on our timescales of interest.  We can therefore rewrite equation (\ref{eqn:phi_induction}) as
\begin{equation}
\frac{\partial B_{\phi h}}{\partial z} \approx - \frac{3}{2} \frac{\Omega_{K}}{v_{zh}} \left( B_{rh} + \frac{h}{r} B_{zh} \right) -  B_{\phi h} \frac{\partial \ln v_{zh}}{\partial z} - \frac{B_{\phi h}}{r} \left[ \frac{v_{rh}}{v_{zh}} \frac{\partial \ln \left( v_{rh} B_{\phi h} \right)}{\partial \ln r} \right] .
\label{eqn:phi_induction_stationary}
\end{equation}
We can simplify this equation by assuming that $\rho v_{z}$ is constant with height so that there is no net buildup of mass in the vertical outflow.  This is a reasonable assumption for thin disks for the timescales of interest.  Thus, $\partial \ln v_{zh} / \partial z = - \partial \ln \rho_{h} / \partial z$.  Numerical simulations show that this quantity is, in turn, well-estimated by $\sim H^{-1}$ throughout the atmosphere of the disk, even in time-dependent systems in which radiation and magnetic fields help to provide vertical support against gravity \citep{Hirose2006,Krolik2007}.  Equation (\ref{eqn:phi_induction_stationary}) then becomes
\begin{equation}
\frac{\partial B_{\phi h}}{\partial z} \sim - \frac{\Omega_{K}}{v_{zh}} \left( B_{rh} + \frac{h}{r} B_{zh} \right) - \frac{B_{\phi h}}{H} \left( 1 + \frac{H}{r} \left[ \frac{v_{rh}}{v_{zh}} \frac{\partial \ln \left( v_{rh} B_{\phi h} \right)}{\partial \ln r} \right] \right) .
\label{eqn:phi_induction_stationary_estimate}
\end{equation}
If we combine this equation with the definition of $H_{B}$, use ${\bf \nabla \cdot B} = 0$, and then ignore terms of order $\sim H/r$ (assuming that the logarithmic radial derivatives are of order unity, which should be true except at boundary regions where the magnetic field structure changes dramatically), we obtain
\begin{equation}
H_{B} \sim H \frac{f_{B}}{1-f_{B}} ,
\label{eqn:hb}
\end{equation}
where $f_{B} \equiv \left( v_{zh}/H \Omega_{K} \right) \left( - B_{\phi h} / B_{rh} \right)$ is typically a positive number in any magnetic field geometry (due to Keplerian shear).  We therefore see that unless $v_{zh}$ approaches the sound speed, we have $H_{B} \lesssim H$ for typical generic field geometries.  This corresponds to our approximation in the main body of the paper.  Furthermore, since $v_{zh}$ is simply the initial speed at which material is launched off the disk surface (before it enters any jet acceleration region), we do not expect it to be large; a reasonable estimate might be $v_{zh} \sim \left< v_{r} \right>$, where $\left< v_{r} \right> \sim \alpha \left( H / r \right) H \Omega_{K}$ is the speed at which material in the main body of the disk accretes inward.  In that case, we would have $H_{B} \ll H$, which is even more favorable for inward advection.

The above analysis shows that the vertical magnetic stress potentially available at $z=h$ in an accretion disk is more than enough to drag the surface layer inward.  In fact, if we consider equation (\ref{eqn:drag_same_speed}) in light of these results, the immediate suggestion is that the surface layer can advect inward much {\it faster} than the main body of the disk.  The problem with this conclusion, however, is that the above analysis only considered the atmosphere on its own, without regard to the main body of the disk below it.  A large shear between the main body of the disk and the surface layer is unlikely to be stable.  Although the physics in the interface between the turbulent body of the disk and the nonturbulent region above it is complicated, we can estimate the effects of a large vertical shear in $v_{r}$ in the case of ideal MHD.  We use the $r$ component of equation (\ref{eqn:induction_equation}), which is
\begin{equation}
\frac{\partial B_{rh}}{\partial t} = \frac{\partial}{\partial z} \left( v_{r} B_{z} - v_{z} B_{r} \right)_{h} ,
\end{equation}
so that there is a term $B_{zh} \partial v_{rh} / \partial z$ which tends to produce $B_{rh}$ on a typical timescale $\left( \partial v_{rh} / \partial z \right)^{-1}$.  Thus a large vertical shear in $v_{r}$ will rapidly change the radial magnetic field---and thereby the vertical stress in accordance with equation (\ref{eqn:phi_induction_stationary_estimate})---in the appropriate direction for the shear to be reduced.  This suggests that very small values of $H_{B}$ are not sustainable, and while it is difficult to predict the exact behavior, we expect that in the general case the base of the nonturbulent surface layer will advect inward at a similar speed as the main body of the disk.  Note that turbulent stresses may also play a role in this region in ensuring that the disk and surface layer advect inward together (similar to a ``friction'' term).  We do not consider their effect here except to note that they are probably smaller than the large-scale magnetic stresses.  (Recall that the large-scale magnetic energy density is by definition comparable to the thermal pressure at $z=h$, while the turbulent vertical stresses are likely to be smaller by a factor of $\sim \alpha$.)\\

\end{appendix}

\end{document}